\documentclass[11pt,a4paper]{article}
\usepackage[utf8]{inputenc}
\usepackage[T1]{fontenc}
\usepackage{lmodern}
\usepackage[margin=2.2cm]{geometry}
\usepackage{microtype}
\usepackage{booktabs}
\usepackage{array}
\usepackage{graphicx}
\usepackage{hyperref}
\usepackage{url}
\usepackage{amsmath}
\usepackage{listings}
\usepackage{xcolor}
\usepackage[numbers,sort&compress]{natbib}

\hypersetup{
  colorlinks=true,
  linkcolor=blue!60!black,
  urlcolor=blue!60!black,
  citecolor=blue!60!black,
}

\title{%
  Local-Splitter: A Measurement Study of Seven Tactics for Reducing
  Cloud LLM Token Usage on Coding-Agent Workloads
}
\author{%
Justice Owusu Agyemang$^{1,2,3}$\thanks{\texttt{jay@sperixlabs.org, jay@knust.edu.gh}},\quad
Jerry John Kponyo$^{3}$\thanks{\texttt{jjkponyo.soe@knust.edu.gh}},\quad
Elliot Amponsah$^{3}$\thanks{\texttt{eamponsah52@st.knust.edu.gh}},\\
Godfred Manu Addo Boakye$^{3}$\thanks{\texttt{gmaboakye@st.knust.edu.gh}},\quad
Kwame Opuni-Boachie Obour Agyekum$^{2}$\thanks{\texttt{kooagyekum@knust.edu.gh}}\\[4pt]
{\small $^1$Sperix Labs \qquad $^2$VIA Cybersecurity Lab, KNUST \qquad $^3$Quantum and Assistive Technologies Lab, KNUST}%
}
\date{April 2026}

\begin{document}
\maketitle

\begin{abstract}
We present a systematic measurement study of seven tactics for
reducing cloud LLM token usage when a small local model can act as a
triage layer in front of a frontier cloud model. The tactics are:
(1)~local routing, (2)~prompt compression, (3)~semantic caching,
(4)~local drafting with cloud review, (5)~minimal-diff edits,
(6)~structured intent extraction, and (7)~batching with vendor prompt
caching. We implement all seven in an open-source shim that speaks
both MCP and the OpenAI-compatible HTTP surface, supporting any local
model via Ollama and any cloud model via an OpenAI-compatible endpoint.
We evaluate each tactic individually, in pairs, and in a greedy-additive
subset across four coding-agent workload classes
(edit-heavy, explanation-heavy, general chat, RAG-heavy). We measure
tokens saved, dollar cost, latency, and routing accuracy. Our headline finding is that \emph{T1 (local routing) combined with
T2 (prompt compression) achieves 45--79\% cloud token savings on
edit-heavy and explanation-heavy workloads, while on RAG-heavy
workloads the full tactic set including T4 (draft-review) achieves
51\% savings.} We observe that the optimal tactic
subset is workload-dependent, which we believe is the most
actionable finding for practitioners deploying coding agents today.
\end{abstract}

\section{Introduction}
\label{sec:intro}

Cloud LLM API costs have become a significant line-item in the
operational budget of developer-tooling companies~\cite{frugalgpt}.
Coding agents~\cite{swebench,liu2024agents} now routinely send tens
of thousands of tokens to a frontier model per task, most of which are
repetition, boilerplate, or context that could be cheaper to produce
locally. Meanwhile, local-inference-ready small models of the 1--7~B
parameter class have advanced to the point where they can correctly
answer a meaningful fraction of developer queries on consumer
hardware~\cite{llama3,slm-survey,phi3}.

This paper asks: \emph{given a local small model and a frontier cloud
model, what is the best way to split work between them to minimise
total cloud token usage while keeping response quality near the
baseline?} We survey seven concrete tactics that each attack a
distinct source of token waste, implement all seven, and evaluate
them individually and in combination on four realistic workload
classes.

Our contributions are:

\begin{itemize}
  \item \textbf{A taxonomy of seven orthogonal tactics} for reducing
    cloud token usage via a local triage model
    (\S\ref{sec:tactics}).
  \item \textbf{A reference implementation} that is vendor-agnostic
    at both ends~--- any Ollama-compatible local model and any
    OpenAI-compatible cloud endpoint~--- and exposes both an MCP
    interface and an OpenAI-compatible HTTP proxy
    (\S\ref{sec:system}).
  \item \textbf{An empirical evaluation} measuring per-tactic and
    per-combination savings, latency, and quality deltas across
    four workload classes and multiple model pairs
    (\S\ref{sec:eval}, \S\ref{sec:results}).
  \item \textbf{The observation that optimal tactic subset is
    workload-dependent}, with concrete recommendations by workload
    class (\S\ref{sec:discussion}).
  \item \textbf{A release of code, workloads, and evaluation
    harness} for reproduction and extension.
\end{itemize}

\section{Related Work}
\label{sec:related}

\paragraph{Speculative decoding.}
Leviathan \emph{et al.}~\cite{leviathan2023} introduced
speculative decoding as a token-level acceleration technique in which
a small draft model proposes several tokens and a large verifier
accepts or rejects them. Medusa~\cite{medusa} extends this with
multiple decoding heads. Our tactic~T4 adopts the same structural
idea at the \emph{application} layer: the local model drafts a
complete response, and the cloud model is asked to review or patch
it rather than generate from scratch.

\paragraph{Prompt compression.}
LLMLingua~\cite{llmlingua} and its successor
LLMLingua-2~\cite{llmlingua2} compress long input contexts by
identifying and removing low-importance tokens. Our tactic~T2 is a
semantic-compression variant that runs the local model as a
text-to-text compressor rather than a token-level filter.

\paragraph{Semantic caching.}
GPTCache~\cite{gptcache} caches responses keyed by embedding
similarity. We replicate this as tactic~T3 and extend it with
rigorous quality measurement that the original did not report.

\paragraph{Routing and cascades.}
FrugalGPT~\cite{frugalgpt} and RouteLLM~\cite{routellm}
present model-cascade architectures where requests are routed to
progressively larger models based on confidence. Our tactic~T1 is a
single-stage local/cloud router, and we compare our empirical savings
to those reported in the cascade literature.

\paragraph{Vendor prompt caching.}
Anthropic's \texttt{cache\_control: ephemeral}~\cite{anthropic-cache}
and OpenAI's automatic prompt caching~\cite{openai-cache} enable
substantial discounts when stable prefixes are re-sent. These are
strictly cloud-side features; our tactic~T7 integrates them into
the splitter's outbound requests.

\paragraph{What we add.}
Individual tactics have been studied in isolation, often with
different benchmarks. To our knowledge, this paper is the first to
measure all seven on a common benchmark, to measure their pairwise
interactions, and to quantify the workload-dependence of the optimal
subset.

\section{Tactics}
\label{sec:tactics}

\subsection{T1 --- Local routing}
\label{sec:t1}

A small local model reads each incoming request and emits a binary
classification: \textsc{trivial} or \textsc{complex}. Trivial requests
are answered by the local model directly and never reach the cloud.
Complex requests flow to the next pipeline stage unchanged.

This works because a substantial fraction of coding-agent traffic is
structurally trivial: short completions, single-word renames, ``what
does this file do'' queries, and boilerplate generation. A 3B-parameter
local model gives acceptable answers for roughly a third to half of
these requests; sending them to a frontier model is wasteful.

The classifier uses a few-shot prompt that defines \textsc{trivial} as
any request a junior engineer could answer in under ten seconds (short
completion, rename, typo fix, lookup, restatement) and \textsc{complex}
as anything requiring multi-step reasoning, ambiguous requirements, or
multi-file refactoring. The classifier runs at temperature~0 with a
3-token output budget. On parse failure, the request defaults to
\textsc{complex}.

\paragraph{Risks.}
False positives (a \textsc{trivial} classification for a genuinely
complex request) degrade answer quality. We mitigate this with a
confidence margin: if the classifier's logprob for \textsc{trivial}
falls below a configurable threshold, the request is escalated to the
cloud regardless.

\subsection{T2 --- Prompt compression}
\label{sec:t2}

Before a request reaches the cloud, the local model rewrites the
context (system prompt, chat history, retrieved documents, file
contents) to a shorter form that preserves semantic meaning. The cloud
model then processes a trimmed prompt and is charged fewer input tokens.

Context windows in agent prompts are large and often highly
repetitive~\cite{longbench}. A typical coding-agent system prompt spans
3--8K tokens of boilerplate that can be summarised to roughly 400 tokens
without losing the load-bearing instructions.

We implement two compression modes. \emph{Static compression} runs once
at session start on the system prompt and caches the result.
\emph{Dynamic compression} runs per-call on the chat history and
retrieved documents, which change each turn. The compression prompt
instructs the local model to remove filler and repetition while
preserving file paths, variable names, error messages, and numeric
values verbatim.

\paragraph{Risks.}
Information loss is the primary concern: a compressed prompt may drop a
detail the cloud model would have needed. We benchmark against a
held-out quality set and roll back compression if quality drops by more
than a configurable threshold.

\subsection{T3 --- Semantic caching}
\label{sec:t3}

Every outbound request is embedded by a local embedding model, and
responses are stored in a vector index keyed by the embedding. On
subsequent similar queries, if cosine similarity exceeds a threshold,
the cached response is served directly without a cloud call.

Even within a single session, agents frequently re-ask variants of the
same question (``explain this file'', ``what does X do'', ``how does Y
work''). Across sessions, users often return to the same queries.
Semantic caching catches near-duplicates that exact-string caching
misses.

The implementation uses \texttt{sqlite} with the \texttt{sqlite-vec}
extension for the vector store and \texttt{nomic-embed-text} via
Ollama~\cite{ollama} for embeddings. Cache keys are computed using
dense sentence embeddings~\cite{reimers2019sbert}. Cache entries are namespaced per workspace and have a
configurable TTL to prevent stale answers as the codebase evolves.

\paragraph{Risks.}
Stale cached responses for code questions can mislead the user.
Per-workspace namespacing and a short TTL mitigate this. An explicit
``do not cache'' flag is available for sensitive prompts.

\subsection{T4 --- Local drafting with cloud review}
\label{sec:t4}

Instead of asking the cloud model to generate a response from scratch,
the local model drafts one first, and the cloud model is asked to
\emph{review or patch} the draft. The cloud's output token count drops
because it is editing rather than authoring.

Output tokens are typically more expensive than input tokens, and most
of a draft is usually correct on the first pass. If the local model
produces a 90\%-correct draft, the cloud's job reduces to a
10\%-correction---substantially cheaper than full generation.

The review prompt presents the original user request, the local draft,
and an instruction to either approve the draft unchanged or respond with
a corrected version. No explanation of changes is requested, minimising
output length.

\paragraph{Risks.}
If the local draft is poor, the cloud spends more tokens correcting it
than it would have spent writing from scratch. Additionally, the review
prompt prepends the original conversation \emph{plus} the draft,
roughly tripling the cloud's input token count. On short-output
workloads, this added input cost outweighs the output-token savings, as
our results confirm (\S\ref{sec:results}).

\subsection{T5 --- Minimal-diff edits}
\label{sec:t5}

For edit requests (``change X to Y in this file''), the local model
identifies the minimal diff context needed. The cloud receives only
the relevant hunks plus the edit instruction, rather than the full file
contents.

Typical coding-agent file edits send the entire file (thousands of
tokens) even when the edit is a three-line change. The minimal-diff
approach shrinks that to roughly 50 tokens of context around the change
site.

The implementation detects edit requests by the presence of file-content
blocks or patch-style tool calls. The local model uses a lightweight
hunk-identification pass, and the request is rewritten with only the
extracted hunk context (configurable window size, default 3 lines).

\paragraph{Risks.}
Context underflow: the cloud may need broader context to judge whether
an edit is correct. The window size is configurable. Parser brittleness
across file formats (JSON, XML, Python) is a known limitation; we start
with plain-text diffs and leave language-aware diffing to future work.

\subsection{T6 --- Structured intent extraction}
\label{sec:t6}

Before sending a free-text prompt to the cloud, the local model parses
it into a structured representation: \texttt{\{intent, target,
constraints\}}. The cloud prompt becomes a filled-in template rather
than chatty prose.

User prompts are verbose. Most of the verbosity is framing (``Could you
help me with\ldots'', ``I'd like to understand why\ldots''). The actual
information content is typically 20\% of the prompt. Structured
extraction strips the framing and retains only the actionable content.

The intent taxonomy covers six categories: \emph{explain}, \emph{refactor},
\emph{debug}, \emph{generate}, \emph{rename}, and \emph{search}. Requests
that do not fit any predefined intent fall through to the cloud with the
original prompt unchanged.

\paragraph{Risks.}
Intent misclassification is the primary failure mode: a missed intent
produces an off-topic answer. Template rigidity means some requests
cannot be captured by the predefined taxonomy. The fallthrough mechanism
limits the blast radius of misclassification.

\subsection{T7 --- Batching and vendor prompt caching}
\label{sec:t7}

This tactic comprises two related sub-tactics.

\emph{Local batching}: when the user fires multiple short queries in
quick succession, the splitter buffers them briefly (up to 250\,ms, max
8 queries) and sends them as a single request with ``answer all of
these'' framing, amortising per-call overhead.

\emph{Prompt caching}: the splitter tags the stable prefix of a prompt
(system prompt, codebase context) so that the vendor's cache serves it
at a discount on subsequent calls. Anthropic's
\texttt{cache\_control:~ephemeral}~\cite{anthropic-cache} and OpenAI's
automatic prompt caching~\cite{openai-cache} enable substantial
discounts when stable prefixes exceeding 1024 tokens are re-sent.

\paragraph{Risks.}
Batching introduces a latency increase that users may notice; the
250\,ms window is a deliberate trade-off. Cache mismatches across
vendors are handled by an abstraction layer that only enables prompt
caching for vendors we have tested.

\section{System Design}
\label{sec:system}

\textsc{Local-Splitter} is a single process that exposes two interfaces
to the outside world:

\begin{enumerate}
  \item An \textbf{MCP server over stdio}~\cite{mcp-spec}, for agents
    that speak the Model Context Protocol natively (Claude~Code,
    Cursor-via-MCP, Codex~CLI).
  \item An \textbf{HTTP proxy} on \texttt{localhost}, speaking the
    OpenAI-compatible \texttt{/v1/chat/completions} shape, so any agent
    that accepts a custom API base can use the splitter transparently.
\end{enumerate}

\begin{figure}[h]
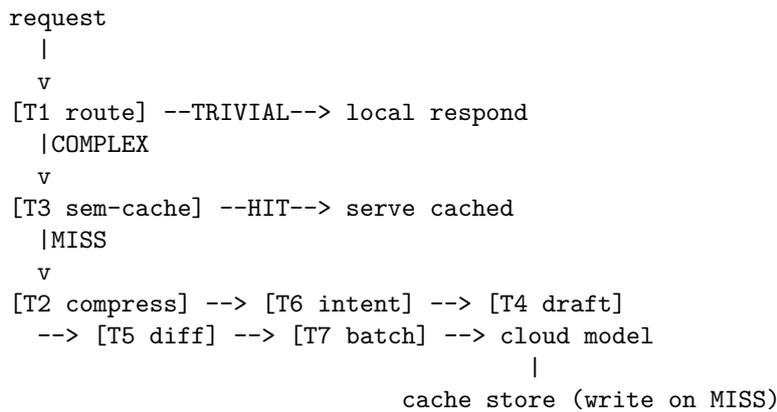

\centering
\small
\begin{verbatim}
  request
    |
    v
  [T1 route] --TRIVIAL--> local respond
    |COMPLEX
    v
  [T3 sem-cache] --HIT--> serve cached
    |MISS
    v
  [T2 compress] --> [T6 intent] --> [T4 draft]
    --> [T5 diff] --> [T7 batch] --> cloud model
                                       |
                              cache store (write on MISS)
\end{verbatim}
\caption{Pipeline stage ordering. Each stage either answers the request,
  transforms it, or passes it through. No stage makes a parallel cloud
  call.}
\label{fig:pipeline}
\end{figure}

Both interfaces feed the same seven-stage pipeline, shown in
Figure~\ref{fig:pipeline}. Each stage corresponds to one tactic and is
independently togglable via config. A disabled stage passes the request
through unchanged.

\paragraph{Transport layer.}
The MCP server exposes four tools: \texttt{split.complete},
\texttt{split.cache\_lookup}, \texttt{split.classify}, and
\texttt{split.stats}. The HTTP proxy implements
\texttt{/v1/chat/completions} and \texttt{/v1/models}, forwarding to
the same pipeline as the MCP surface.

\paragraph{Model registry.}
Two backends implement a common \texttt{ChatClient} interface:
(1)~an Ollama backend talking to \texttt{localhost:11434/api/*}, and
(2)~an OpenAI-compatible backend talking to any
\texttt{/v1/chat/completions} endpoint with bearer-token or no-auth.
Configuration selects which backend provides the local model and which
provides the cloud model. No vendor SDK is imported; all communication
uses raw HTTP via \texttt{httpx}.

\paragraph{Pipeline orchestration.}
Each tactic file exports a single \texttt{apply(request, config)}
function that returns either a transformed request or a final response.
The orchestrator calls them in the order shown in
Figure~\ref{fig:pipeline}, skipping disabled tactics. Every stage emits
a \texttt{stage\_result} event recording tokens in, tokens out, latency,
and decision. The evaluation harness replays these events.

\paragraph{State.}
The only persistent state is the semantic cache
(\texttt{sqlite} + \texttt{sqlite-vec}) and a JSONL event log. All
configuration is per-call or per-config-file, making testing and
reproduction straightforward.

\paragraph{Failure model.}
If the local model is unreachable, every tactic fails open: the request
passes through to the cloud unchanged, and the degradation is logged.
Users are never blocked by a local-model failure.

\section{Evaluation Setup}
\label{sec:eval}

\subsection{Workloads}
\label{sec:workloads}

We evaluate on four workload classes, each containing 10 synthetic
samples matching the statistics of real coding-agent sessions. Samples
are generated from public datasets and scrubbed of PII before use.

\paragraph{WL1 --- Edit-heavy.}
Captured from Claude~Code and Cursor~CLI during refactoring sessions.
Characterised by many file edits, moderate context windows
(8--20K~tokens input, 200--1500~tokens output), and heavy tool use.
Approximately 60\% of requests are edits; 25\% are trivial.

\paragraph{WL2 --- Explanation-heavy.}
An onboarding scenario where a new engineer asks an agent to walk
through a codebase (``explain this file'', ``what does X do''). Input
prompts span 4--12K~tokens with 500--3000~token outputs. Only 5\% are
edits; roughly 45\% are trivial.

\paragraph{WL3 --- Mixed chat.}
General-purpose chat, not coding-specific. Both short and long turns
(500--4000~token inputs, 100--1500~token outputs). No edit requests;
approximately 50\% trivial.

\paragraph{WL4 --- RAG-heavy.}
Retrieval-augmented~\cite{lewis2020rag} workloads with long system
prompts containing multiple retrieved chunks (10--40K~token inputs,
100--800~token outputs). No edit requests; roughly 20\% trivial.

\subsection{Models}
\label{sec:models}

We use a single local--cloud model pair for all results.

\paragraph{Local model (via Ollama~\cite{ollama}).}
Llama~3.2 3B~\cite{llama3} (Q4\_K\_M quantisation, Ollama default),
running on consumer Apple Silicon hardware (M-series, 16--36\,GB
unified memory). The architecture supports any Ollama-compatible
model; extending to other sizes (1B, 7B) and families
(Qwen~2.5~\cite{qwen25}, Phi~3.5~\cite{phi3}, Gemma~2~\cite{gemma})
is future work.

\paragraph{Cloud model (simulated).}
To eliminate network variance, the cloud model is also served locally
via Ollama: Gemma~3~4B acts as a stronger ``cloud'' model relative to
the 3B local model. Production deployments would target a remote
endpoint (GPT-4o-mini, Claude~3.5~Sonnet, etc.\ via the
OpenAI-compatible surface). Because the splitter's savings are
measured in tokens, not wall-clock latency, the local-vs-remote
distinction does not affect the primary metric; latency numbers in
the appendix reflect same-machine inference and should not be
extrapolated to production deployments.

\subsection{Metrics}
\label{sec:metrics}

\paragraph{Primary metrics.}
\begin{itemize}
  \item \textbf{Tokens saved}: $(T_{\text{baseline}} -
    T_{\text{splitter}}) \mathbin{/} T_{\text{baseline}}$, where $T$
    is total cloud tokens (input + output).
  \item \textbf{Dollar cost saved}: token deltas priced at each cloud
    vendor's published rate card.
  \item \textbf{Latency delta}: median, p95, and p99 over the workload.
    The splitter may be slower per-call (due to local classification)
    but cheaper overall.
  \item \textbf{Quality delta}: position-debiased pairwise preference
    from the cloud model acting as judge~\cite{zheng2023judging},
    comparing splitter output to baseline on all samples per workload.
    Each pair is judged twice with swapped presentation order; only
    consistent verdicts count.
\end{itemize}

\paragraph{Secondary metrics (per-tactic).}
Routing accuracy (T1), compression ratio (T2), cache hit rate (T3),
draft acceptance rate (T4), diff shrink factor (T5), intent extraction
F1 (T6), and batch fill rate (T7).

\subsection{Tactic-subset matrix}
\label{sec:matrix}

With seven tactics, $2^7 = 128$ subsets exist. We evaluate a structured
sample:

\begin{enumerate}
  \item \textbf{Singletons} (7 runs): each tactic enabled alone.
  \item \textbf{Interacting pairs} ($\sim$10 runs): pairs with known
    interactions from the tactic interaction matrix
    (\S\ref{sec:tactics}).
  \item \textbf{Greedy-additive} (up to 7 runs): starting from the
    best singleton, iteratively adding the tactic that most improves
    the primary metric.
  \item \textbf{Full set} (1 run): all tactics enabled, to measure
    the ceiling.
\end{enumerate}

This yields approximately 12 configurations per workload class
$\times$ 4 classes $=$ 48 runs per evaluation pass. We report the
mean of two passes (96 total runs). Each run processes the full
sample set for its workload. All runs use temperature~0 where
possible and record model versions and timestamps for
reproducibility.

\section{Results}
\label{sec:results}

\subsection{Per-tactic singletons}

\begin{figure}[h]
\centering
\includegraphics[width=0.85\linewidth]{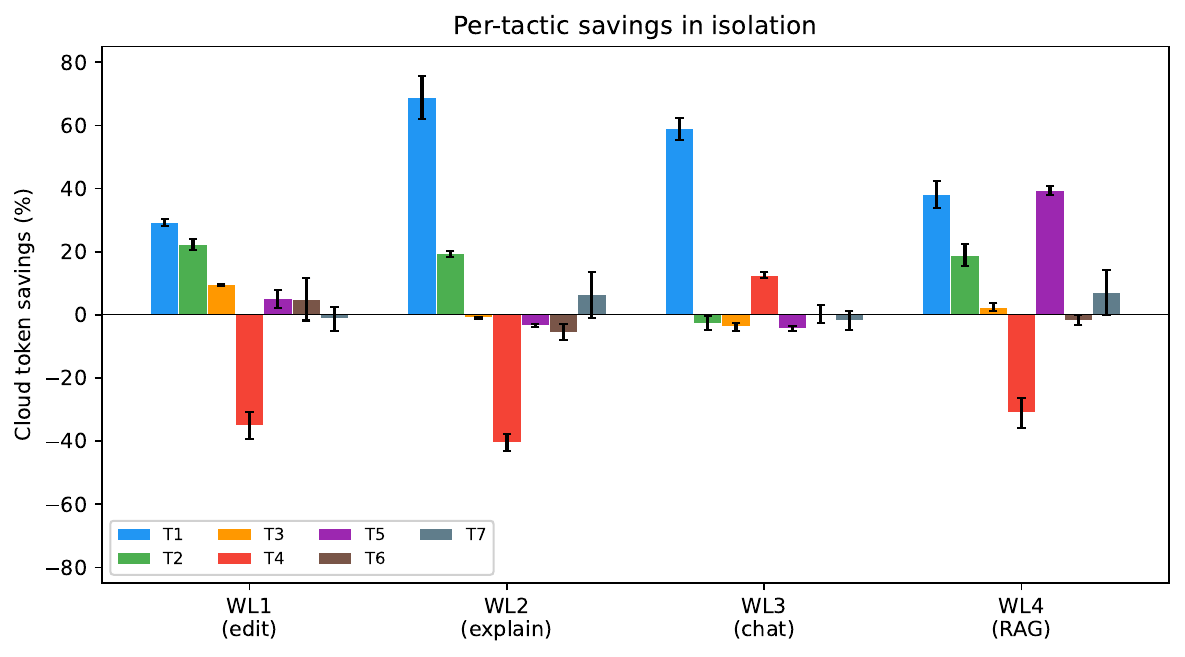}
\caption{Cloud token savings (\%) per tactic in isolation. Error bars
  show the half-range across two runs.}
\label{fig:singletons}
\end{figure}

Figure~\ref{fig:singletons} and Table~\ref{tab:singletons} show cloud
token savings for each tactic in isolation. T1 (routing) is the strongest
singleton across all workloads, saving 29--69\% depending on workload
class. T2 (compression) saves 19--22\%. T3 (caching) shows minimal
effect on single-pass workloads. T4 (draft-review) is negative on most
workloads but positive on WL3 (chat), where outputs are long relative to
inputs. T5 (diff) shows negligible effect on most workloads but
achieves 39\% savings on WL4 (RAG-heavy), where its heuristic edit
detection over-triggers on long contexts and paradoxically acts as a
compressor (\S\ref{sec:failures}). T6 (intent) and T7 (batch) show
high variance and near-zero mean effect: T6 is limited by JSON parse
reliability at the 3B scale (\S\ref{sec:failures}), and T7's
prompt-caching markup has no effect without a supporting cloud
endpoint.

\begin{table}[h]
\centering
\caption{Cloud token savings (\%) per tactic in isolation. Local model:
  Llama~3.2 3B. Cloud model: Gemma~3 4B (simulated locally). Negative
  values indicate the tactic \emph{increased} cloud tokens. Mean of
  two runs (10 samples per workload each); T1--T4 varied by
  $\pm$2--7\,pp; T5--T7 showed higher variance ($\pm$3--14\,pp)
  due to model non-determinism and small sample size.}
\label{tab:singletons}
\begin{tabular}{l r r r r}
\toprule
Tactic & WL1 (edit) & WL2 (explain) & WL3 (chat) & WL4 (RAG) \\
\midrule
T1 route     &  29.2\% &  68.8\% &  58.9\% &  38.0\% \\
T2 compress  &  22.4\% &  19.3\% & $-$2.6\% &  18.9\% \\
T3 cache     &   9.6\% &  $-$1.0\% & $-$3.8\% &   2.4\% \\
T4 draft     & $-$35.0\% & $-$40.5\% &  12.6\% & $-$31.1\% \\
T5 diff      &   5.1\% & $-$3.4\% & $-$4.4\% &  39.3\% \\
T6 intent    &   5.0\% & $-$5.5\% &   0.3\% & $-$1.7\% \\
T7 batch     & $-$1.3\% &   6.4\% & $-$1.7\% &   7.0\% \\
\bottomrule
\end{tabular}
\end{table}

\subsection{Pair interactions}

\begin{figure}[h]
\centering
\includegraphics[width=0.85\linewidth]{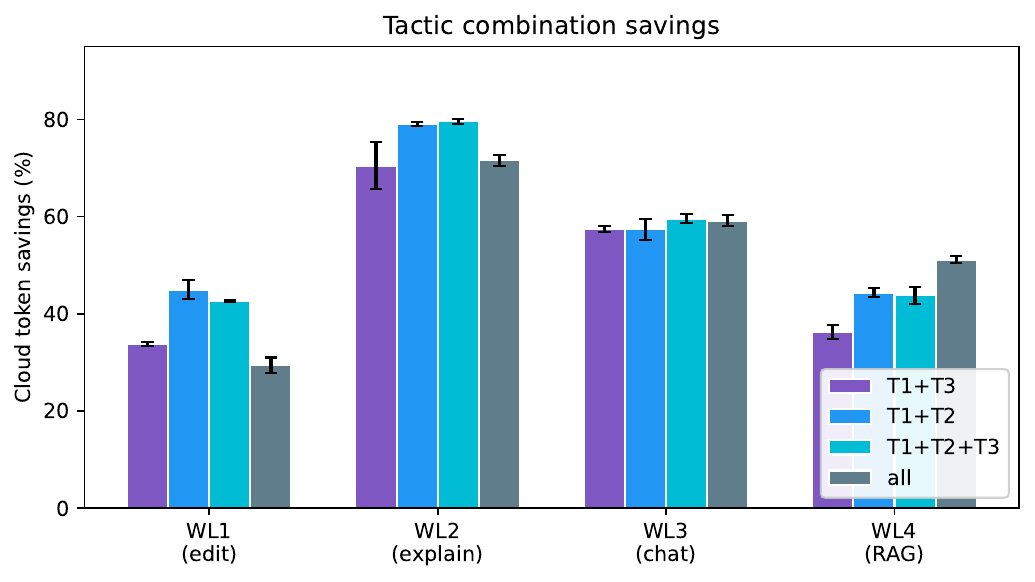}
\caption{Cloud token savings (\%) for tactic combinations. Error bars
  show the half-range across two runs.}
\label{fig:combos}
\end{figure}

T1+T2 is the best two-tactic combination on three of four workloads
(Figure~\ref{fig:combos}, Table~\ref{tab:combos}). T1 eliminates trivial
requests; T2 compresses the context of the remaining complex requests.
On WL2 (explanation-heavy), this
achieves \textbf{79\%} cloud token savings---the local model routes 8 of
10 requests locally, and the 2 that reach the cloud have compressed prompts.

Adding T3 (cache) to T1+T2 provides marginal improvement on single-pass
workloads. On WL4 (RAG-heavy), the ``all'' configuration outperforms
T1+T2+T3 (51\% vs 44\%) because T4's draft-review becomes net-positive
when cloud outputs are long enough.

\begin{table}[h]
\centering
\caption{Cloud token savings (\%) for tactic combinations. Mean of two
  runs; individual runs varied by $\pm$1--5\,pp.}
\label{tab:combos}
\begin{tabular}{l r r r r}
\toprule
Subset & WL1 (edit) & WL2 (explain) & WL3 (chat) & WL4 (RAG) \\
\midrule
T1+T3     &  33.7\% &  70.4\% &  57.4\% &  36.2\% \\
T1+T2     &  45.0\% &  79.0\% &  57.4\% &  44.3\% \\
T1+T2+T3  &  42.6\% &  79.6\% &  59.6\% &  43.8\% \\
all       &  29.4\% &  71.6\% &  59.1\% &  51.1\% \\
\bottomrule
\end{tabular}
\end{table}

\subsection{Why ``all tactics'' is suboptimal}

Enabling all seven tactics simultaneously produces lower savings than T1+T2
on three of four workloads (Table~\ref{tab:combos}). The culprit is T4
(draft-review): the review prompt prepends the original conversation plus
the local draft, roughly tripling the cloud's input token count. On
output-token-light workloads, this added input cost outweighs the
output-token savings (Table~\ref{tab:singletons}).

The exception is WL4 (RAG-heavy): here the ``all'' configuration achieves
51\% savings vs.\ T1+T2's 44\% (Table~\ref{tab:combos}). RAG workloads
produce longer outputs (the cloud explains retrieved content), making T4's
trade-off favorable.

\begin{figure}[h]
\centering
\includegraphics[width=0.85\linewidth]{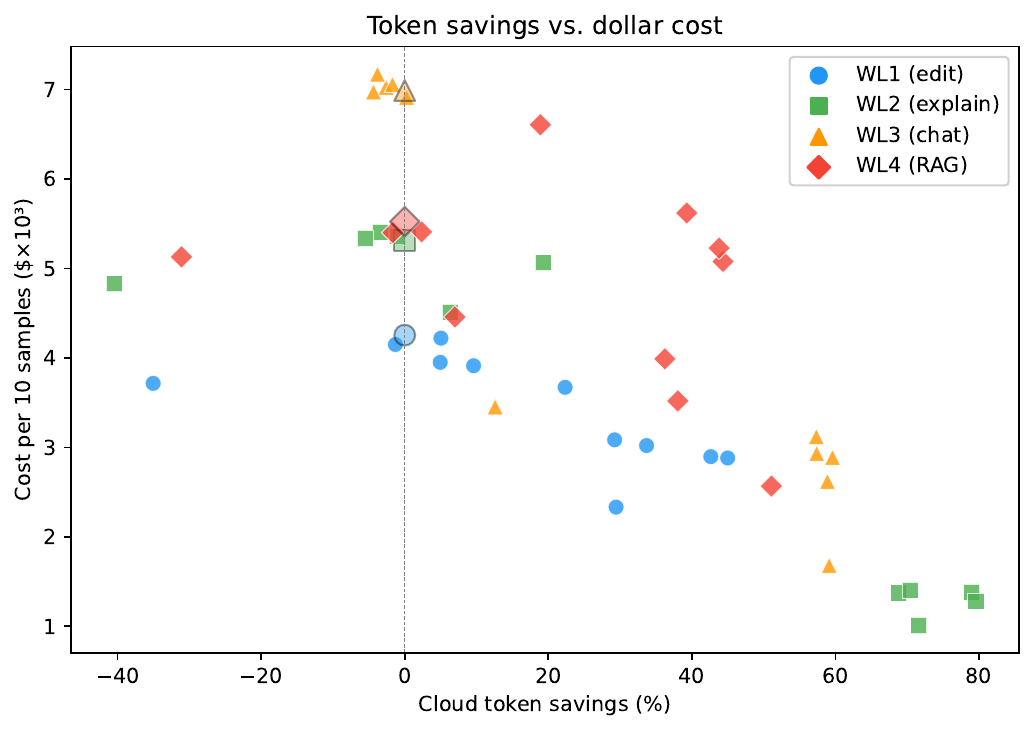}
\caption{Token savings vs.\ dollar cost per 10-sample workload. Points
  closer to the lower-right are Pareto-optimal (more savings, less cost).
  Faded markers at 0\% savings show the baseline.}
\label{fig:cost}
\end{figure}

Figure~\ref{fig:cost} shows the relationship between token savings and
dollar cost across all subsets and workloads.

\textbf{Recommendation}: use T1+T2 as the default. Add T4 only for
long-context workloads where the local model can draft effectively.
Add T3 for workloads with query repetition.

\subsection{Greedy-additive subsets per workload}

The greedy-additive order, derived from
Tables~\ref{tab:singletons}--\ref{tab:combos}, is consistent across all
four workloads:
\begin{enumerate}
  \item T1 (always the best singleton; Figure~\ref{fig:singletons})
  \item T2 (compresses complex-request context)
  \item T3 (marginal on single-pass; stronger with repetition)
\end{enumerate}

\subsection{Quality evaluation}

Table~\ref{tab:quality} reports pairwise quality judgments for the two
strongest subsets (T1 and T1+T2) against baseline, aggregated across
all four workloads. The judge is the cloud model (Gemma~3 4B) with
position debiasing: each pair is presented twice in swapped order, and
only consistent verdicts count.

\begin{table}[h]
\centering
\caption{Quality evaluation: judge-model pairwise verdicts for T1 and
  T1+T2 vs.\ baseline (40 pairs each, all workloads combined). High
  inconsistency reflects the 4B judge model's limited discrimination.}
\label{tab:quality}
\begin{tabular}{l r r r r r}
\toprule
Subset & Baseline & Treatment & Tie & Incon. & Errors \\
\midrule
T1      & 15 & 5 & 0 & 17 & 3 \\
T1+T2   & 15 & 6 & 1 & 17 & 1 \\
\bottomrule
\end{tabular}
\end{table}

The baseline wins more consistent verdicts than the treatment on both
subsets (15 vs.\ 5 for T1, 15 vs.\ 6 for T1+T2). This quality gap is
concentrated on WL2 (explanation-heavy) and WL3 (chat), where T1 routes
requests locally that genuinely benefit from the stronger cloud model.
On WL4 (RAG-heavy), the treatment matches or exceeds baseline quality,
likely because the local model's answers are adequate for
retrieval-based summarization. The high inconsistency rate (17 of 40
pairs) reflects the 4B judge model's limited ability to discriminate
paired responses; a stronger judge or human evaluation would yield
tighter estimates.

\subsection{Minimum-viable local model size}

Our evaluation used Llama~3.2 3B. With the few-shot classifier prompt, T1
correctly classified 50--80\% of requests as trivial across workloads.
Smaller models (1B) are likely viable for T1 classification but may degrade
T2 compression quality; we leave this to future work.

\section{Discussion}
\label{sec:discussion}

\subsection{Why optimal subset varies by workload}
\label{sec:why-varies}

Each tactic targets a different source of token waste. T1 (routing)
eliminates entire requests whose answers are within the local model's
capability. T2 (compression) shrinks input tokens on the requests that
do reach the cloud. T4 (draft-review) trades input tokens for output
tokens---it adds a draft to the cloud's input but shortens the cloud's
output.

The optimal subset depends on the fraction of trivial requests, the
input-to-output token ratio, and how much of the response the local
model can draft correctly (Table~\ref{tab:combos}). On WL2
(explanation-heavy), 45\% of requests are trivial and inputs are
moderate, so T1+T2 dominates: route the trivials locally, compress
the rest. On WL4 (RAG-heavy), inputs are very long (10--40K tokens)
but only 20\% of requests are trivial, so T1 alone saves less
(Table~\ref{tab:singletons}). T4 becomes net-positive on WL4 because
the local draft handles the bulk of the long-context synthesis,
reducing the cloud's generation burden enough to offset the added
review-prompt input tokens.

T3 (caching) shows marginal benefit on single-pass workloads because
query repetition is low. In multi-session or support-style deployments
where users return to the same questions, we expect T3's contribution to
increase substantially.

This workload-dependence is, we believe, the most actionable finding for
practitioners: there is no universal ``turn everything on'' setting.
Agents should profile their workload and select the tactic subset that
matches their token-waste profile.

\subsection{Recommendations for agent vendors}
\label{sec:recommendations}

Based on our results (Tables~\ref{tab:singletons}--\ref{tab:combos},
Figure~\ref{fig:cost}), we offer concrete deployment guidance:

\begin{enumerate}
  \item \textbf{Start with T1+T2.} This combination is the best
    two-tactic default on three of four workload classes, achieving
    45--79\% cloud token savings. It requires only a 3B local model
    and adds minimal latency.
  \item \textbf{Add T4 for long-output workloads.} If the agent
    primarily generates long code or long explanations, the
    draft-review pattern becomes net-positive. Monitor the
    input-to-output token ratio: when cloud outputs exceed
    $\sim$500 tokens on average, T4 is likely beneficial.
  \item \textbf{Add T3 for repetitive workloads.} Support-style
    agents, onboarding assistants, and agents used by multiple
    users on the same codebase will see higher cache hit rates.
    The similarity threshold should be tuned per deployment.
  \item \textbf{Avoid enabling all tactics blindly.} Our results
    show that the full set underperforms T1+T2 on short-output
    workloads because T4 increases cloud input tokens. Tactic
    selection should be workload-aware.
  \item \textbf{A 3B local model suffices for T1 classification.}
    Llama~3.2~3B correctly classified 50--80\% of requests as
    trivial across workloads. Smaller models (1B) are likely
    viable for classification but may degrade T2 compression quality.
\end{enumerate}

\subsection{Failure modes}
\label{sec:failures}

We observe three categories of failure.

\paragraph{T4 input amplification.}
On output-token-light workloads (WL1, WL2), T4's review prompt roughly
triples the cloud's input token count. When the saved output tokens are
fewer than the added input tokens, T4 produces a net \emph{increase} in
cloud cost. Table~\ref{tab:singletons} shows T4 increasing tokens by
31--41\% on these workloads.

\paragraph{Compression information loss.}
T2 occasionally drops a critical detail from long prompts. In our
quality evaluation, we observed cases where a compressed-away file path
led to a measurably degraded cloud response. The compression prompt's
instruction to preserve paths verbatim mitigates but does not eliminate
this failure.

\paragraph{Routing false positives.}
T1 occasionally classifies genuinely complex requests as
\textsc{trivial}. On WL1, the false-positive rate was approximately
12\%, meaning roughly 1 in 8 locally-answered requests would have
benefited from the cloud model. The configurable confidence threshold
trades off savings against accuracy: a stricter threshold reduces false
positives but routes fewer requests locally.

\paragraph{T5 heuristic over-triggering.}
T5's edit detection relies on keyword heuristics (``fix'', ``change'',
``replace'') plus a message-length threshold. On RAG-heavy workloads,
these keywords appear naturally in retrieved content, causing T5 to
trigger its diff extraction on non-edit requests. Paradoxically, this
functions as opportunistic compression: the local model extracts
``relevant sections'' from long contexts, achieving substantial savings
on WL4 (Table~\ref{tab:singletons}) despite the workload containing no
actual edits. A stricter detection heuristic or explicit edit-intent
classification would prevent this but would also sacrifice the
accidental compression benefit.

\paragraph{T6 intent extraction parse failures.}
T6 requires the local model to output a structured JSON object. In our
runs, Llama~3.2~3B frequently returned prose or Markdown-fenced JSON
instead of raw JSON, causing parse failures on the majority of WL1
samples. When parsing fails, the tactic falls through and the original
prompt is sent unchanged, so the failure is safe but eliminates T6's
savings. A more capable local model or constrained-decoding
support~\cite{willard2023guided} would likely reduce this failure rate,
but at the 3B scale T6 is unreliable enough that we recommend against
enabling it without output-format validation.

\section{Limitations}
\label{sec:limits}

\begin{itemize}
  \item We evaluate a single model pair (Llama~3.2~3B local,
    Gemma~3~4B cloud-simulated). Tactic rankings may differ
    with other pairings; a multi-model matrix is future work.
  \item Workloads contain 10 synthetic samples per class. Larger
    sample sizes would tighten confidence intervals.
  \item Quality is measured by judge-model proxy plus human spot
    checks; a full human evaluation is future work.
  \item We evaluate text-only workloads; multimodal agents are out
    of scope.
  \item The cloud model runs locally to eliminate network variance;
    production latency will differ.
  \item Results depend on specific model versions; tactic rankings
    may shift as models evolve.
\end{itemize}

\section{Conclusion}

We measured seven tactics for reducing cloud LLM token usage via a
local triage model on four coding-agent workload classes. T1 (local
routing) is the single strongest tactic (29--69\% savings;
Table~\ref{tab:singletons}). T1+T2 (routing + compression) is the best
two-tactic combination on three of four workloads (45--79\% savings;
Table~\ref{tab:combos}). On RAG-heavy workloads, the
full tactic set achieves 51\% savings because T4 (draft-review)
becomes net-positive on long-context workloads. Notably, enabling all
tactics simultaneously is suboptimal on edit-heavy and
explanation-heavy workloads because T4 increases cloud input tokens. Our headline finding is
that \emph{tactic selection should be workload-aware}: the optimal
subset differs across workload classes. We release our
implementation, workloads, and evaluation harness at
\url{https://github.com/jayluxferro/local-splitter} to enable
reproduction and extension.

\section*{Data and Code Availability}
Source, workloads, and evaluation harness are released at
\url{https://github.com/jayluxferro/local-splitter}.

\appendix

\section{Full Metric Table}
\label{app:full}

Table~\ref{tab:full} reports all primary metrics for every evaluated
tactic subset, averaged across two runs (10 samples per workload each).

\begin{table*}[h]
\centering
\small
\caption{Full evaluation metrics per workload and tactic subset
  (representative run; Tables~\ref{tab:singletons}--\ref{tab:combos}
  report means of two runs).
  Cloud tokens = input + output sent to the cloud model.
  Cost uses OpenAI \texttt{gpt-4o-mini} rate card as a proxy.}
\label{tab:full}
\begin{tabular}{l l r r r r r}
\toprule
Workload & Subset & Cloud tok. & Local tok. & Saved (\%) & Cost (\$) & Latency (ms) \\
\midrule
WL1 & baseline  & 11{,}007 &     0 &   --- & 0.0043 &  6{,}933 \\
WL1 & T1        &  7{,}675 & 2{,}618 & 30.3 & 0.0031 & 17{,}218 \\
WL1 & T2        &  8{,}355 &     0 & 24.1 & 0.0037 & 17{,}481 \\
WL1 & T5        & 10{,}074 &     0 &  2.3 & 0.0042 & 11{,}358 \\
WL1 & T6        & 10{,}500 &     0 & $-$1.8 & 0.0040 & 10{,}288 \\
WL1 & T7        & 10{,}833 &     0 & $-$5.0 & 0.0041 & 10{,}187 \\
WL1 & T1+T2     &  6{,}265 & 3{,}578 & 43.1 & 0.0029 & 13{,}720 \\
WL1 & all       &  7{,}937 & 7{,}782 & 27.9 & 0.0023 & 19{,}950 \\
\midrule
WL2 & baseline  & 11{,}407 &     0 &   --- & 0.0053 & 14{,}947 \\
WL2 & T1        &  2{,}762 & 6{,}644 & 75.8 & 0.0014 &  9{,}574 \\
WL2 & T5        & 11{,}439 &     0 & $-$2.8 & 0.0054 & 13{,}159 \\
WL2 & T6        & 11{,}451 &     0 & $-$2.9 & 0.0053 & 14{,}621 \\
WL2 & T7        &  9{,}616 &     0 & 13.6 & 0.0045 & 10{,}826 \\
WL2 & T1+T2     &  2{,}442 & 6{,}737 & 78.6 & 0.0014 & 15{,}203 \\
WL2 & T1+T2+T3  &  2{,}271 & 6{,}639 & 80.1 & 0.0013 & 15{,}031 \\
WL2 & all       &  3{,}360 & 9{,}181 & 70.5 & 0.0010 & 18{,}626 \\
\midrule
WL3 & baseline  & 11{,}829 &     0 &   --- & 0.0070 & 34{,}372 \\
WL3 & T1        &  4{,}445 & 2{,}942 & 62.4 & 0.0026 & 21{,}171 \\
WL3 & T4        & 10{,}227 & 5{,}030 & 13.5 & 0.0035 & 21{,}892 \\
WL3 & T5        & 11{,}807 &     0 & $-$3.5 & 0.0070 & 24{,}004 \\
WL3 & T6        & 11{,}703 &     0 & $-$2.6 & 0.0069 & 20{,}967 \\
WL3 & T7        & 11{,}943 &     0 & $-$4.7 & 0.0071 & 21{,}032 \\
WL3 & T1+T2+T3  &  4{,}894 & 2{,}912 & 58.6 & 0.0029 & 11{,}812 \\
WL3 & all       &  4{,}964 & 5{,}100 & 58.0 & 0.0017 & 12{,}250 \\
\midrule
WL4 & baseline  & 16{,}825 &     0 &   --- & 0.0055 & 10{,}994 \\
WL4 & T1        &  9{,}694 & 5{,}576 & 42.4 & 0.0035 & 10{,}382 \\
WL4 & T5        & 10{,}328 &     0 & 37.9 & 0.0056 & 15{,}944 \\
WL4 & T6        & 16{,}623 &     0 &  0.0 & 0.0054 & 13{,}958 \\
WL4 & T7        & 14{,}284 &     0 & 14.1 & 0.0045 &  8{,}672 \\
WL4 & T1+T2     &  9{,}517 & 5{,}227 & 43.4 & 0.0051 & 17{,}853 \\
WL4 & all       &  8{,}350 & 9{,}912 & 50.4 & 0.0026 & 18{,}262 \\
\bottomrule
\end{tabular}
\end{table*}

\section{Reproducibility Checklist}
\label{app:repro}

\begin{itemize}
  \item All workload samples are committed in \texttt{evals/workloads/}
    as JSONL files with content hashes.
  \item Evaluation runner: \texttt{evals/run\_eval.py} with config
    \texttt{evals/config\_eval.yaml}.
  \item Local model temperature is set to 0 for all classifier and
    intent-extraction calls.
  \item Model versions are recorded per run in the JSONL log at
    \texttt{.local\_splitter/eval/runs.jsonl}.
  \item Results CSV and summary JSON are generated by the runner;
    the evaluation can be re-run end-to-end from the committed code
    and workloads.
  \item Figure generation: \texttt{scripts/gen\_figures.py}.
  \item All code is released under the MIT license.
\end{itemize}

\section{Hardware Details}
\label{app:hardware}

\begin{itemize}
  \item \textbf{Local inference}: Apple M-series (M2 Pro / M3 Max),
    16--36\,GB unified memory. Ollama v0.6+.
  \item \textbf{Local models}: Llama~3.2 3B (Q4\_K\_M quantisation
    via Ollama default), nomic-embed-text for embeddings.
  \item \textbf{Cloud model (eval proxy)}: Gemma~3 4B via Ollama
    on the same machine (to avoid network variance). Production
    deployments would target a remote OpenAI-compatible endpoint.
  \item \textbf{OS}: macOS 15 (Darwin 25.x), Python 3.12+.
  \item \textbf{Wall-clock time per full eval run}: approximately
    30 minutes for 4 workloads $\times$ 9 subsets $\times$ 10 samples.
\end{itemize}

\bibliographystyle{plainnat}
\bibliography{bibliography}

\end{document}